# Spontaneous dehydrogenation of methanol over defect-free MgO(100) thin film deposited on molybdenum

Zhenjun Song, Hu Xu*

Department of Physics, South University of Science and Technology of China, Shenzhen, 518055, China

**Abstract:**

The dehydrogenation reaction of methanol on metal supported MgO(100) films has been studied by employing periodic density functional calculations. As far as we know, the dehydrogenation of single methanol molecule over inert oxide insulators such as MgO has never been realized before without the introduction of defects and low coordinated atoms. By depositing the very thin oxide films on Mo substrate we have successfully obtained the dissociative state of methanol. The dehydrogenation reaction is energetically exothermic and nearly barrierless. The metal supported thin oxide films studied here provide a versatile approach to enhance the activity and properties of oxides.

## Introduction

Metal-oxide nanostructures, which possess unique physicochemical properties, have great potential for device applications, including transparent electrodes, high-mobility transistors, gas sensors, photonic devices, energy harvesting and storage devices, and nonvolatile memories.[1] Much work has been devoted extensively to the potential applications of metal oxides in heterogeneous catalysis.[2,3] MgO, one of the most important model system to investigate oxide surface chemistry, has been focused on for its very simple rock-salt crystalline structure and valuable catalytic properties. The electronic structure of MgO only involves s and p electrons, which can be considered an ideal system in order to study the catalytic properties of more complex solids. MgO itself can serves as an effective catalyst for oxidation and photochemical reactions.[4-7] Once deposited on metal substrate, the electronic structure and energy levels of MgO can be controlled by different lattice parameters in the epitaxial film and the interface bonding.[8]

Adapting the inherent characteristic of the substrate, the surface orientation, and the thickness of the supported oxide films is especially promising for synthesizing artificial materials with new properties.[9] The chemical reactivity of insulating MgO film is considerably enhanced by deposition on transition metal substrates.[10-15] Although most first-principles calculations on MgO-metal interfaces are mostly confined to Ag metal support, the refractory transition metal Mo which sustains high annealing temperatures can make the supported MgO thin films of better quality and smaller roughness.[8] In sensor technology and catalysis, organic molecules adsorbing onto dielectric substrates and oxides are key building blocks.[16-18]

Methanol, the simplest aliphatic alcohol, is one of the most common laboratory organic solvent. Methanol can be used for dissolving mineral salt, coating material, pigment, alkaloid, and acetyl cellulose. Using solar-generated hydrogen, methanol could be produced from direct reduction of $CO_2$ in heterogeneously catalyzed processes,[19] where the methanol become a sustainable source of liquid fuel. Molecular-level understanding the adsorption behavior and chemical bonding of organic solvents on oxide surface is of particular importance for enhancing catalytic performance of oxides toward organic reactions. As far as we know, the spontaneous dissociative adsorption properties of alcohol on insulator surface such as MgO(100) has never been revealed. In this paper, the dissociative adsorption of methanol over molybdenum supported MgO(100) films (denoted as MgO(100)/Mo(001)) is studied using density functional calculations.

## Methods

Periodic density functional calculations have been performed by using Vienna *ab initio* simulation package (VASP)[20, 21] to determine all structural, energetic and electronic results. Perdew-Burke-Ernzerhof (PBE) functional[22] within generalized gradient approximation (GGA) to describe exchange and correlation effects, which includes an accurate description of the uniform electron gas, correct behavior under uniform scaling and a smooth potential. Projector Augmented Wave (PAW)[20, 23] technique is used to describe electronic structure and treat the interactions between valence electrons and the core. Electron configurations $1s^1$, $2s^22p^2$, $2s^22p^4$, $2s^2$ and $4d^55s^1$ are used to describe valence electrons in H, C, O, Mg and Mo atoms. The Kohn-Sham orbitals were expanded by using plane waves with a kinetic energy cutoff of 500 eV. Through spin-polarized plane wave calculations with a k-point mesh of 9×9×9, the lattice constants of the Mo bulks are determined to be 3.151 Å which is in good agreement to reported experimental values 3.14 Å.[24] I use a four atomic layer Mo slab with the two bottom layers fixed at bulk positions while the other two metal layers and the MgO film are fully relaxed until all atomic Hellmann-Feynman forces are less than 0.02 eV/Å. I found that an even larger number of Mo layers do not change the surface chemical properties of the MgO films. Convergence criterion for energy minimization is $1.0\times10^{-5}$ eV. For the calculations, we use supercells containing 16Mg + 16O atoms per layer or 16Mo atoms per layer. Gamma-centered k-point meshes 2×2×1 and 4×4×1 is used to sample the first Brillouin zone, for structure optimization and energy calculation respectively.

In all calculations the periodically repeated slabs are separated by a thick vacuum larger than 17 Å. The energy barriers and transition states are located by using the climbing image nudged elastic band (CI-NEB) method[25], which is an efficient method for searching the minimum energy path (MEP) connecting the given initial and final states. Because the highest-energy image is trying to maximize its energy along the band and minimize energy in all other directions, the exact saddle point along the reaction path is easier to find. Therefore, less number of intermediate images is needed in CI-NEB than NEB.

## Results and discussion

### Methanol adsorption on MgO(100) surface

The representative binding sites for methanol oxygen on the MgO surface including on top of an O atom, on top of an Mg atom, and above the center of the square formed by nearest-neighbor Mg and O sites, are considered to determine the structural characteristic of adsorption structure. The most stable orientation of methanol is presented in Figure 2. The adsorption of methanol on the stoichiometric and most stable (001) face of MgO is very weak chemical adsorption, with adsorption energy of –0.46 eV. The interaction between hydroxyl of methanol and surface oxygen anion is hydrogen bond. The methanol oxygen binds primarily with surface Mg cation via electrostatic force. The bond length and surface rumpling of MgO $n$ layer of methanol adsorbed onto the MgO(100) surface are listed in Table 1. It is seen that the surface structure does not alter substantially. The average Mg–O bond length at the adsorption site is 2.15 Å, which is merely 0.03 Å longer than that of bulk MgO(100). Comparing with other surface ionic bonds, the Mgs–O1 distance is the largest, which suggests that the formation of Om–Mgs electrostatic interaction and H1···O1 hydrogen bond broke the Mgs–O1 ionic bond slightly. Comparing with the molecular methanol optimized at the same theoretical level, the Om–H1 distance (1.00 Å) is lengthened by 0.03 Å after adsorption, because of the effective interaction between hydroxyl and surface atoms. However, the C–Om distance (1.43 Å) remain unchanged, indicating the chemical interaction is restricted to the hydroxyl group of methanol. The surface rumpling is attenuated quickly when layer $n$ is increased, indicating the chemical interaction acting on the MgO(100) film is restricted to the surface layer ($n = 1$, $\Delta z_n = 0.005$ Å).

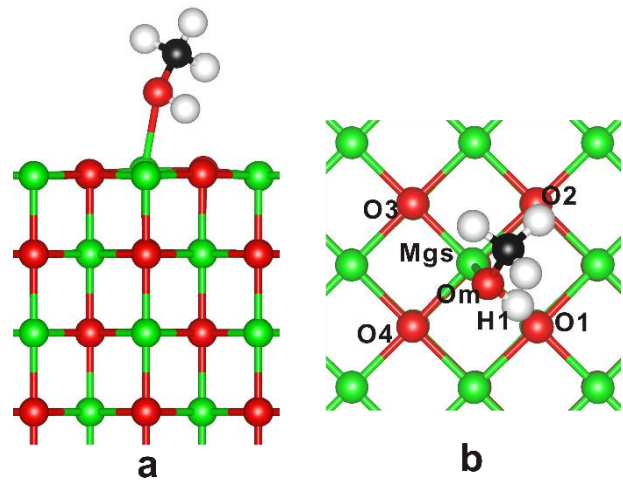

Figure 1. Optimized adsorption geometry of methanol onto the MgO(100) surface: a, side view; b, top view.

Table 1. Calculated bond length and surface rumpling of MgO $n$ layer of methanol adsorbed onto the MgO(100) surface. All units are set in Å.

|  | Bond length |  | Bond length | Layer $n$ | $\Delta z_n$[a] |
|---|---|---|---|---|---|
| C–Om | 1.43 | Mgs–O1 | 2.23 | 1 | 0.005 |
| Om–H1 | 1.00 | Mgs–O2 | 2.15 | 2 | 0.001 |
| Om–Mgs | 2.21 | Mgs–O3 | 2.08 | 3 | 0.001 |
| H1···O1 | 1.82 | Mgs–O4 | 2.13 | 4 | 0.000 |

[a]$\Delta z_n$ is surface rumpling of surface layer $n$, defined as max($z_O - z_{Mg}$). Positive $\Delta z_n$ values correspond to protrusion of oxygen from the layer $n$.

**Methanol adsorption and dissociation on MgO(100)/Mo(001) surfaces**

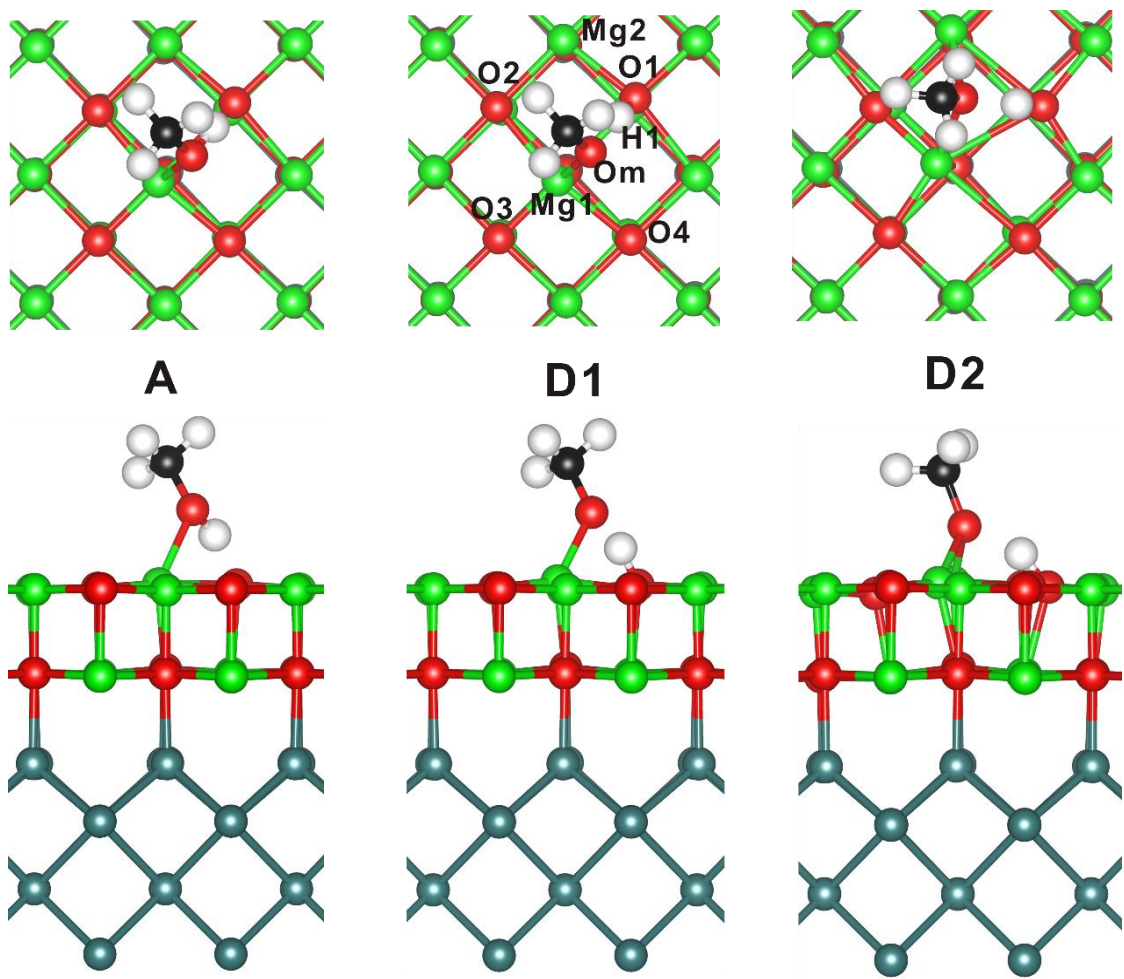

A           D1           D2

Figure 2. Optimized adsorption geometry of methanol onto the 2 ML MgO(100)/Mo(001) surface in molecular adsorption state (A), and in dissociative adsorption states (D1 and D2).

Table 2. Selected bond distances of surface and adsorbing species, at molecular adsorption state (A) and dissociative adsorption states (D1 and D2).

| bond | 1 ML | | | 2 ML | | | 3 ML | | |
|---|---|---|---|---|---|---|---|---|---|
| | A | D1 | D2 | A | D1 | D2 | A | D1 | D2 |
| Mg1–O1 | 2.72 | 2.98 | 3.02 | 2.70 | 2.81 | 2.94 | 2.60 | 2.76 | 2.92 |
| Mg1–O2 | 2.27 | 2.30 | 2.74 | 2.29 | 2.36 | 2.01 | 2.31 | 2.43 | 2.06 |
| Mg1–O3 | 2.09 | 2.05 | 2.10 | 2.07 | 2.04 | 2.08 | 2.08 | 2.05 | 2.07 |
| Mg1–O4 | 2.26 | 2.45 | 2.93 | 2.28 | 2.33 | 2.89 | 2.26 | 2.26 | 2.79 |
| Mg2–O1 | 2.25 | 2.09 | 2.94 | 2.24 | 2.29 | 2.93 | 2.26 | 2.37 | 2.89 |
| Mg2–O2 | 2.45 | 2.84 | 2.71 | 2.35 | 2.34 | 2.74 | 2.29 | 2.27 | 2.58 |
| C–Om | 1.43 | 1.41 | 1.43 | 1.43 | 1.41 | 1.42 | 1.43 | 1.41 | 1.42 |
| Om–H1 | 1.04 | 1.39 | 1.54 | 1.08 | 1.40 | 1.52 | 1.04 | 1.37 | 1.47 |
| Om–Mg1 | 2.08 | 1.95 | 2.12 | 2.08 | 1.97 | 2.21 | 2.10 | 1.98 | 2.17 |
| Om–Mg2 | 3.54 | 3.57 | 2.13 | 3.44 | 3.40 | 2.09 | 3.48 | 3.40 | 2.16 |
| O1–H1 | 1.57 | 1.09 | 1.01 | 1.44 | 1.08 | 1.01 | 1.56 | 1.10 | 1.02 |

Table 3. Adsorption energies ($E_{ad}$) for molecular and dissociative adsorption states (A, D1, and D2), and activation barriers ($E_a$) for obtaining dissociative adsorption states (D1 and D2).

| Surface | $E_{ad}$ | | | $E_a$ | |
|---|---|---|---|---|---|
| | A | D1 | D2 | A→D1 | D1→D2 |
| 1 ML MgO(100)/Mo(001) | –0.71 | –0.82 | –0.97 | 0 | 0.02 |
| 2 ML MgO(100)/Mo(001) | –0.77 | –0.79 | –0.85 | 0 | 0.07 |
| 3 ML MgO(100)/Mo(001) | –0.73 | –0.71 | –0.71 | 0.03 | 0.08 |

The binding sites for methanol oxygen over the MgO surface deposited on Mo substrate including on top of an O atom, on top of an Mg atom, and above the center of the square formed by nearest-neighbor Mg and O sites, are considered to determine the structural characteristic of adsorption and dissociation structure. The most stable orientation of methanol over the Mo supported MgO surface is presented in Figure 2. The nondissociative adsorption of methanol on the stoichiometric MgO is calculated to be chemical adsorption, with adsorption energy of -0.71, -0.77 and -0.73 eV respectively for 1-3 ML films. The interaction between hydroxyl of methanol and surface oxygen anion is hydrogen bond for nondissociative adsorption state, with O1…H1 bond length of 1.57, 1.44, and 1.56 Å for 1-3 ML films. The strongest hydrogen bonding between O1 and H1 in 2 ML film is consistent with the sequence of nondissociative adsorption energies (2 ML film have the most negative adsorption energy -0.77 eV).

For the first dissociation state D1, the O1-H1 distances are calculated to be 1.09, 1.08 and 1.10 Å for 1-3 ML films (as listed in Table 2), conforming the surface hydroxyl formation on metal supported MgO(100). The dissociative adsorption energies of methanol on 1-3 ML MgO/Mo(001) are calculated to be -0.82, -0.79, -0.71 eV respectively, which decreases in absolute value with increasing film thicknesses. The dissociative adsorption energies for 1-2 ML MgO/Mo(001) are more negative than corresponding nondissociative adsorption energies. However, the 3 ML thickness film have dissociative adsorption energy less negative than the corresponding nondissociative adsorption energy, indicating the more thicker oxide film are unfavorable for dissociating methanol molecule. Because of the O1-H1 covalent bond formation, the Mg1-O1 bond is partly broken. Comparing with the nondissociative state, the Mg1-O1 ionic bonds are lengthened by 0.26, 0.11 and 0.16 Å for 1-3 ML films after dissociation (Table 2). The largest detachments of Mg1-O1 (0.26 Å) and Mg1-O4 (0.19 Å) for monolayer MgO(100) suggest the presence of severe surface distortion. However, the Mg1-O3 bonds are slightly strengthened, because of the translation of negative charged O3 towards Mg1. The Mg2 atom is farther away from the adsorption site. The bonding of Mg2 can be seen as another detection parameter of the surface micromorphologic alteration. The Mg2-O1 distances are 2.09, 2.29 and 2.37 Å for 1-3 ML films respectively. The Mg2-O2 distances are 2.84, 2.34 and 2.27 Å for 1-3 ML films respectively. Thus, the monolayer MgO(100) experiences most strong surface destruction, even at the surface position far away from the adsorption site. The enhancement of surface destruction at the monolayer MgO(100) can be ascribed to the very thin surface thickness and the important role of oxide-metal interface structure. Interestingly, the thickest 3 ML film deform more severe than the 2 ML film. This can be attributed to the even-odd alteration of the MgO layer numbers. Different from the odd number oxide layers, the oxide with even number layers have alternative $Mg^{2+}$-$O^{2-}$ ordering perpendicular to the surface plane.

The methoxyl group of the D1 state can translate to the center of the two neighboring Mg atoms, which produces the second dissociative state D2. At state D2, the dissociative adsorption energies of 1-2 ML MgO/Mo(001) films have more negative values than corresponding values at both state A and state D1. For 3 ML MgO/Mo(001), the state D2 are isoenergetic with state D1 and both are slightly higher in energy (by 0.02 eV) than nondissociative adsorption state, demonstrating the unfavorable dissociation behavior at more thicker oxide films. After the transformation to state D2, the O1-H1 bond distances are further shortened to 1.01 - 1.02 Å. Due to the much strengthening of the O1-H1 covalent bond, all the Mg1-O1 distances are lengthened compared with state D1. All the Mg1-O4 distances for 1-3 ML MgO/Mo(001) are lengthened more than 0.5 Å comparing with corresponding values at nondissociative adsorption state, suggesting the effective bonding of methoxyl group with the Mg1 cation at state D2. As the O1-H1 covalent bonds at state D2 are strengthened very much compared with these at state D1, the Mg2-O1 are lengthened to c.a. 2.9 Å. The different reactivity for methanol on MgO/Mo(001) versus on MgO can be attributed to the greater freedom that the thin MgO films has to deform greatly to accommodate the adsorbates and the high reactive products (hydroxyl and methoxyl).

The results of a Bader charge analysis for the adsorbates (methanol, methoxyl, hydrogen, hydroxyl), the MgO films, and the Mo substrates are presented in Table 4. After dissociation of methanol, the methoxyl groups tend to gain more electrons from the metal supported MgO films. As can be seen in Table 4, the Bader charge of H1 increases from state A to D1 and from D1 to D2. Based on the Bader charge analysis, the dissociation of methanol (A to D1) and the transformation from D1 to D2 increases the binding of H1 atom on the metal MgO surface. The net charge of H1O1 group decreases after dissociation and transformation from D1 to D2. Interestingly, the thicker film possesses more negative charge of the surface hydroxyl. This can be attributed to the chemical instinct of surface oxygen of thicker MgO film, which are less affected by the Mo substrate underneath. During the interaction with the metal substrate and the adsorbates, the MgO films are highly oxidized. The monolayer MgO with positive charges +2.009, +2.115, +2.115 is oxidized more seriously than thicker MgO films. During the dissociation processes (A → D1 → D2), the MgO film shift slightly towards more oxidized states. The Mo substrate are all negatively charged. The Mo substrate in 1 ML MgO/Mo(001) accumulates more electrons than that in thicker films. However, the Mo substrate in 2 ML MgO/Mo(001) are less negative than that in 3 ML film. Therefore, the monolayer and odd-number oxide layers are more easily be oxidized.

Table 4. Bader charges of Methoxyl, H1, H1O1, MgO films, and Mo substrates at different adsorption states A, D1, and D2.

|  | 1 ML MgO/Mo(001) | | | 2 ML MgO/Mo(001) | | | 3 ML MgO/Mo(001) | | |
| --- | --- | --- | --- | --- | --- | --- | --- | --- | --- |
|  | A | D1 | D2 | A | D1 | D2 | A | D1 | D2 |
| Methoxyl | -0.687 | -0.770 | -0.813 | -0.723 | -0.789 | -0.819 | -0.702 | -0.782 | -0.814 |
| H1 | +0.611 | +0.613 | +0.619 | +0.605 | +0.605 | +0.625 | +0.607 | +0.609 | +0.625 |
| H1O1 | -0.880 | -0.854 | -0.837 | -0.974 | -0.939 | -0.897 | -0.993 | -0.947 | -0.900 |
| MgO | +2.009 | +2.115 | +2.115 | +1.489 | +1.592 | +1.594 | +1.577 | +1.635 | +1.639 |
| Mo | -1.933 | -1.958 | -1.921 | -1.371 | -1.409 | -1.401 | -1.482 | -1.462 | -1.450 |

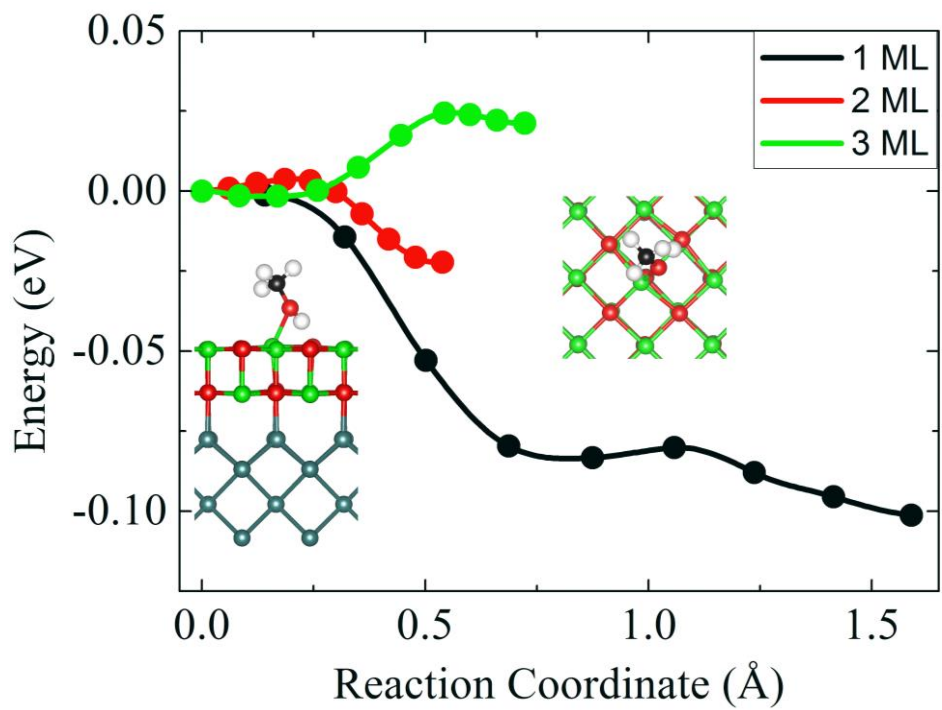

Figure 3. Potential energy profiles for dissociation pathway of chemical adsorbed methanol (A → D1) on 1-3 ML MgO/Mo(100) surfaces. The reaction coordinate is the cumulative distance along the minimum energy path, where the initial position is set to zero.

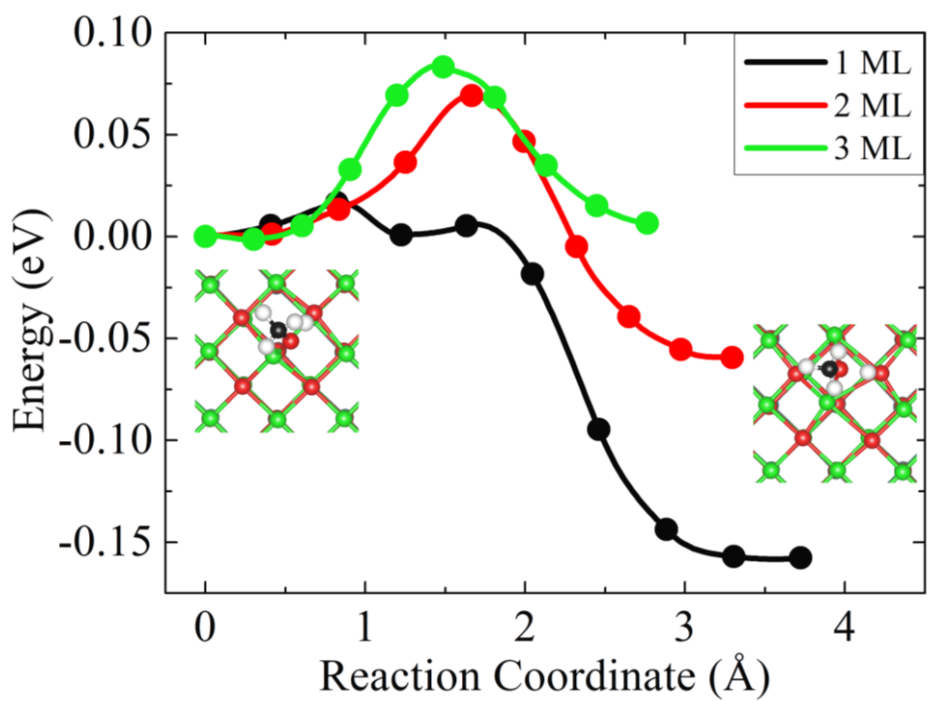

Figure 4. Potential energy profiles for transformation pathway for dissociative methanol (D1 → D2) on 1-3 ML MgO/Mo(100) surfaces. The reaction coordinate is the cumulative distance along the minimum energy path, where the initial position is set to zero.

As illustrated in Figure 3, the minimal energy pathway for methanol dissociation to D1 states are obtained by using CI-NEB method at 0 K. The activation barriers for methanol dissociation are calculated to be 0, 0, 0.03 eV for dissociation process A → D1 on 1-3 ML oxide films respectively, as listed in Table 3. Thus, the dissociation of methanol is barrierless on metal supported very thin MgO films (1-2 ML). Although the usually very strong covalent bonds are being broken, this dehydrogenation process can apparently occur spontaneously on metal supported very thin oxide film. As shown in Figure 4, the transformation from D1 state to D2 state has very small barrier energies, 0.02, 0.07 and 0.08 eV for transformation reaction occurring on 1-3 ML films respectively. The thicker films are slightly unfavorable for the transformation from D1 state to D2 state. The transformation barriers are negligibly small and the two dissociation states D1 and D2 can coexist on MgO surface.

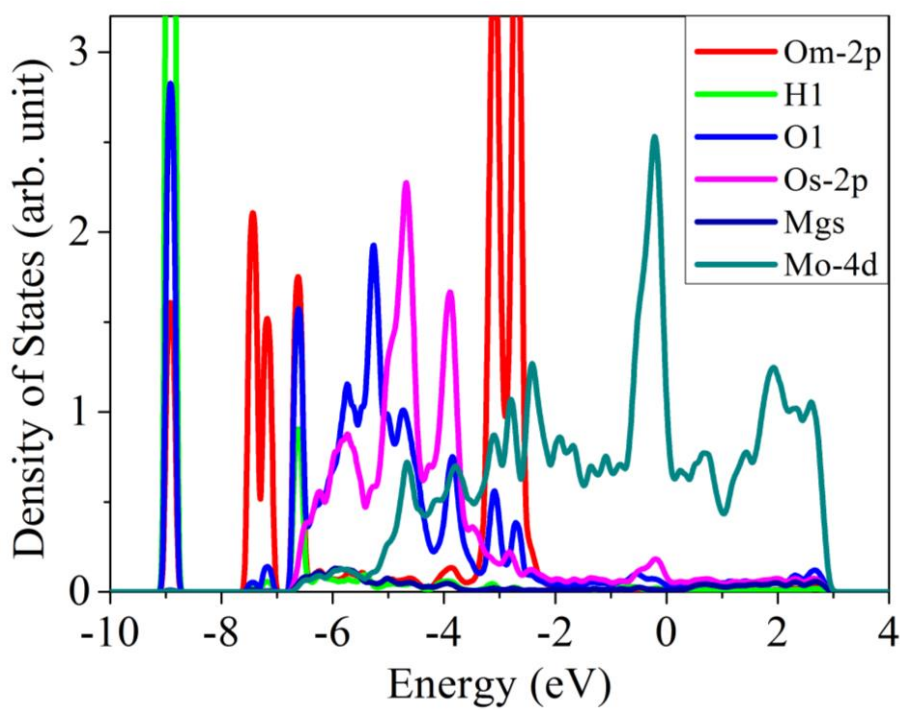

Figure 5. Local density of states (LDOS) for detached H1 atom, the surface oxygen O1 which forms the hydroxyl, and the surface Mg atoms (Mgs), and the projected

density of states (PDOS) for 2p character of surface O atoms (Os-2p), methanol O (Om-2p), and 4d character of interface Mo atoms (Mo-4d) for dissociative state D1 on metal supported oxide film. The Fermi energy is considered as zero energy reference.

The electronic structural properties can improve our understanding of the mechanism for the novel phenomenon of spontaneous methanol dehydrogenation. Here we mainly focus on the electronic structure of methanol adsorbing at metal supported monolayer oxide film, as the monolayer MgO are most favorable for the dehydrogenation reaction. Figure 5 shows the local density of states (LDOS) for detached H1 atom, the surface oxygen O1 which forms the hydroxyl, and the surface Mg atoms (Mgs), and the projected density of states (PDOS) for 2p character of surface O atoms (Os-2p), methanol O (Om-2p), and 4d character of interface Mo atoms (Mo-4d). For comparison, the PDOS for surface O atoms has excluded the contribution form O1 atom. The density of states are averaged to one atom, and the LDOS values for hydrogen are amplified five times for better visibility. From the density of states of surface O and Mg atoms, we can infer that the large band gap of MgO disappeared after supporting on the molybdenum substrate. The modified band structure can be ascribed to the chemical interaction at the interface structure. The density of states of Os-2p overlap largely with Mo-4d, demonstrating that the effective covalent bonding interaction should exist at the interface. As the Om adsorbing at the surface is one part of methoxyl group, which has the molecular characteristic, the 2p bands for methoxyl are very discrete comparing with that of the surface oxygen. The Om-2p band overlaps with H1-1s band near -6.7 eV and -9 eV, proving the very strong hydrogen bonding between Om and H1. The relatively strong band hybridization between H1 and Om-2p agrees well with the partially broken Om-H1 bond with bond length of 1.39 Å. Very different from other surface oxygen atoms, the 2p bands of O1 atom emerge peaks at -9 eV and -6.7 eV. This result further confirm that after methanol dissociation, the hydrogen forms covalent bond with the surface oxygen. The covalent bonding interaction result in the significant lowering of the 2p character of O1 deep into the valence band.

The isosurface of differential charge density is given in Figure 6 (side view and top view). The charge density around bulk Mo atoms accumulate because of the high electron affinity of Mo. Unlike the bulk Mo, at the interface region, all the Mo atoms lose electron to oxygen, and the charge density changes demonstrate the relatively strong covalent bond form between O and Mo. The large gray regions around the methanol adsorption sites demonstrate that the methoxyl group withdraw electrons from metal supported surface. As shown in Figure 6b, the bonding character and strength of ionic O-Mg between two oxide layers undergo considerable change after adsorption and supporting on the transition metal substrate. The ionic bonds has been partially broken because of the electron depletion. This change to very strong ionic bonds may lead to significant changes of surface properties and surface reactions, which has also been verified in our recent study on usual water splitting and hydrogen dissociation over MgO(100)/Mo.[13, 14]

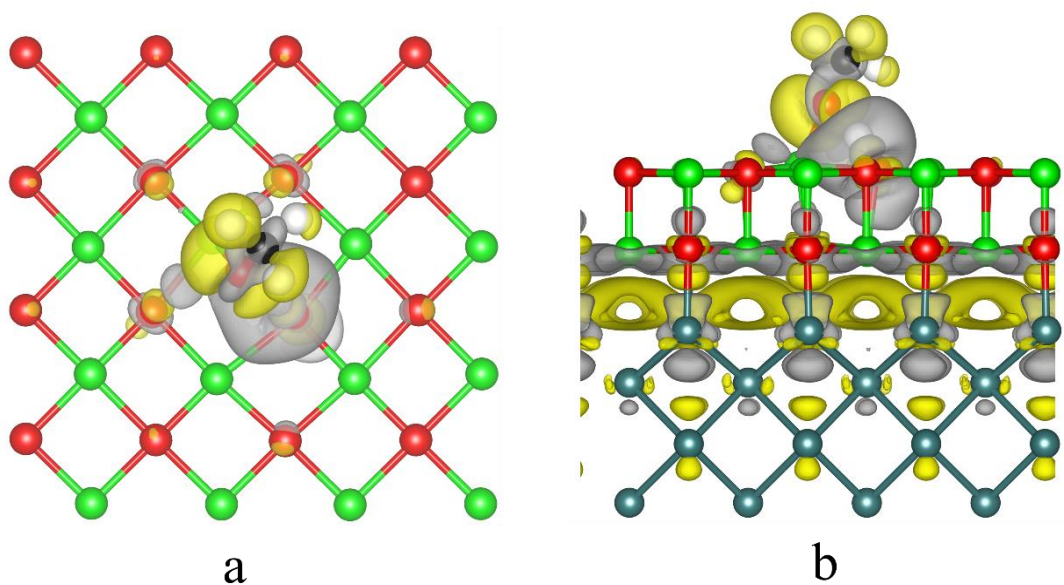

a    b

Figure 6. Top view and side view of differential charge density for methanol dissociation state D1. For clarity, top view shows only electron distribution for the dissociative methanol and the first layer oxide. Differential charge density is defined as $\Delta\rho = \rho(Total) - \rho(Mo) - \rho(MgO) - \rho(methoxyl) - \rho(H)$. The isosurface value is set to be 0.002 e/bohr$^3$. Yellow and gray areas correspond to electron accumulation and electron depletion, respectively.

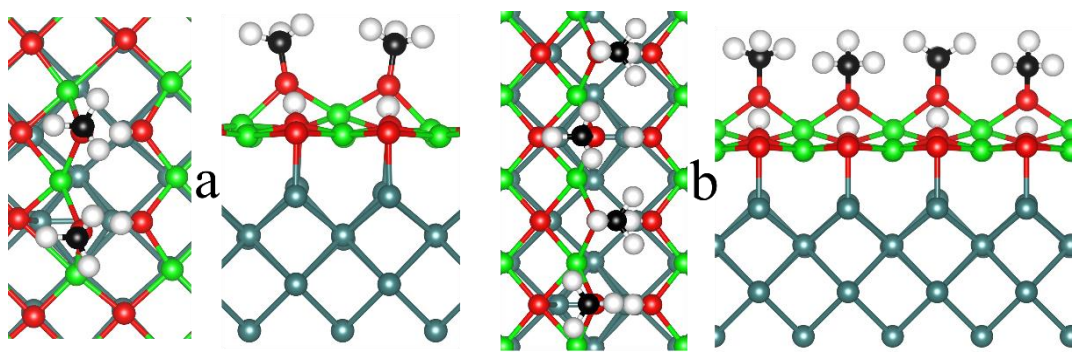

Figure 7. Adsorption geometry of two (a) and four (b) methanol molecules onto the 2 ML MgO(100)/Mo(001) surface in dissociative adsorption states D2.

We further considered the case of dissociative adsorption of methanol molecules with larger coverages, as shown in Figure 7. Two methanol molecules located between three neighboring surface magnesium on 1 ML MgO/Mo(100), can dissociate with energy gain of 2.068 eV. More strikingly, four methanol molecules arranged in a line on 1 ML MgO/Mo(100), can dissociate to form surface hydroxyl and methoxyl groups, which yields even larger energy gain of 4.251 eV. In the latter case, the four methoxyl species are staggered with respect to each other to decrease the space repulsion force. The $O_s$-H all have bond length of ca. 1.01 Å, indicating the definite formation of surface hydroxyl groups. Thus, the dehydrogenation reaction of methanol could occur with larger coverages, when the Van der Waals repulsions between methanol molecules are quite small.

## Conclusion

In conclusion, the dehydrogenation reaction of methanol on metal supported MgO(100) films has been studied systemically by periodic DFT methods. As far as we know, the dehydrogenation of single methanol molecule over inert oxide insulators such as MgO has never been realized before. By depositing the very thin oxide films on Mo substrate we have successfully obtained the dissociative state of methanol. The dehydrogenation reaction is energetically exothermic and barrierless for methanol adsorbing on 1-2 ML MgO(100)/Mo. D1 state can transform to D2 state easily with a very small activation barrier (0.02 eV) and with further energy gain of 0.15 eV for the reaction on 1 ML MgO(100)/Mo. Besides the structural characterization, the electronic properties of the adsorbing species further confirm the formation of methoxyl and surface hydroxyl species. The bonding character and strength of ionic O-Mg bonds undergo considerable change after adsorption and supporting on the transition metal substrate. Because of the electron depletion, the ionic bonds has been partially broken, which leads to significant improvement of surface properties and reactivity. The detailed investigation here opens new perspective for designing new model catalysts to improve the chemical activity of usually inert insulator oxides.